\documentclass[conference]{IEEEtran}
\IEEEoverridecommandlockouts
\usepackage{cite}
\usepackage{amsmath,amssymb,amsfonts}
\usepackage{algorithmic}
\usepackage{graphicx}
\usepackage{textcomp}
\usepackage{xcolor}
\usepackage{listings}
\usepackage{xurl}
\usepackage[hidelinks]{hyperref}
\usepackage{todonotes}[]
\usepackage{booktabs}
\usepackage{tikz}
\usepackage{tcolorbox}
\usepackage{multirow}
\usepackage{centernot}

\makeatletter
\def\footnoterule{\kern-3\p@
  \hrule \@width 2in \kern 2.6\p@} 
\makeatother

\def\BibTeX{{\rm B\kern-.05em{\sc i\kern-.025em b}\kern-.08em
    T\kern-.1667em\lower.7ex\hbox{E}\kern-.125emX}}

\tikzstyle{process} = [rectangle,  rounded corners, minimum width=2cm, minimum height=1cm, text centered, draw=black, fill=gray!30]
\tikzstyle{arrow} = [thick,->,>=stealth]

\newtcolorbox{qoutebox}[3][]
{
  colframe=black!30!white,
  colback  = #2!10,
  #1,
}

\begin{document}

 \title{On the Effect of Instrumentation on Test Flakiness}

\author{
	\IEEEauthorblockN{Shawn Rasheed\IEEEauthorrefmark{1}, Jens Dietrich\IEEEauthorrefmark{2}, Amjed Tahir\IEEEauthorrefmark{3}}
 \IEEEauthorblockA{\IEEEauthorrefmark{1}Universal College of Learning, Palmerston North, New Zealand}
	\IEEEauthorblockA{\IEEEauthorrefmark{2}Victoria University of Wellington, Wellington, New Zealand}
	\IEEEauthorblockA{\IEEEauthorrefmark{3}Massey University, Palmerston North, New Zealand \\  jens.dietrich@vuw.ac.nz; s.rasheed@ucol.ac.nz; a.tahir@massey.ac.nz}
}

\maketitle

\begin{abstract}

Test flakiness is a problem that affects testing and processes that rely on it. Several factors cause or influence the flakiness of test outcomes. Test execution order, randomness and concurrency are some of the more common and well-studied causes. Some studies mention code instrumentation as a factor that causes or affects test flakiness. However, evidence for this issue is scarce. In this study, we attempt to systematically collect evidence for the effects of instrumentation on test flakiness. We experiment with common types of instrumentation for Java programs\textemdash{}namely, application performance monitoring, coverage and profiling instrumentation. We then study the effects of instrumentation on a set of nine programs obtained from an existing dataset used to study test flakiness, consisting of popular GitHub projects written in Java. We observe cases where real-world instrumentation causes flakiness in a program. However, this effect is rare. We also discuss a related issue\textemdash{}how instrumentation may interfere with flakiness detection and prevention.

\end{abstract}


\section{Introduction}
\label{sec:introduction}

A test can only provide useful feedback if it consistently has the same outcome (either pass or fail) for every execution with the same code version. Flaky tests may pass in some runs and fail on others. Test flakiness has been gaining the attention of both academia and industry because of its negative impact on testing, testing-dependent processes (especially automated testing in CI/CD pipelines), and techniques that rely on executing tests \cite{parry2021survey,rasheed2022test}. Several factors can cause test flakiness, including concurrency, test order dependency, network, shared state and platform dependencies. Most of these causes are common across programming languages and platforms \cite{hashemi2022empirical,luo2014empirical,costa2022test}.

One potential factor that may have an impact on test flakiness is instrumentation.
Ideally, for common use cases of instrumentation, such as code coverage, the effect of instrumentation should be transparent to the application. Concerning test flakiness, this means that test outcomes should remain the same with or without instrumentation. This may well not be the case in practice, though. 
A study on code coverage at Google \cite{ivankoviFSE2019} describes flakiness as a cause for failed coverage computation. Lam et al. \cite{wing2019rootcause} explain how their instrumentation for root-causing flakiness interferes with program behaviour and leads to increased/decreased test flakiness.

However, there are gaps in these studies with respect to the question we are interested in. Those studies are not focused on the effect of instrumentation on flakiness, and the full datasets are not available as the studies are from the industry. This work is an attempt towards addressing this, in which we discuss the effect of instrumentation on test flakiness, and how it affects flaky test prevention/detection techniques.
We propose evaluation metrics and perform a preliminary evaluation on a dataset used in a previous test flakiness study to determine whether instrumentation impacts flakiness. In this work, we address the research question: \\\textbf{RQ:} Does instrumentation  increase/decrease test flakiness?

\section{Related work}

\subsection{Flaky test detection}
Several techniques have been proposed to detect flaky tests, most of which determine flakiness by observing transitions of test outcome across multiple runs \cite{bell2018deflaker,dutta2020detecting}. There are two main approaches used for detecting flaky tests: static techniques that rely only on analyzing the test code without actually executing tests or dynamic techniques that involve the execution of tests \cite{rasheed2022test}.
Most of the existing tools focus on specific causes of test flakiness, such as concurrency (e.g., Shaker \cite{silva2020shake}) and test order-dependency (e.g., iFixFlakies \cite{shi2019ifixflakies}). There have been recent attempts to build lightweight static approaches for flaky test prediction that aim to avoid or minimize test reruns \cite{alshammari2021flakeflagger,fatima2022flakify}.

\subsection{Instrumentation and flakiness}
Wing et al. \cite{wing2019rootcause} used instrumentation to record runtime properties for root-causing flaky tests. They reported that their instrumentation could change runtime behaviour and thus decrease or increase test flakiness. In their study, runtime overhead from instrumentation was observed to affect the reproducibility of flaky tests (by executing a random sample from 59 flaky tests, two tests were flaky only with instrumentation and three were flaky without instrumentation). Ivankovi\'{c} et al. \cite{ivankoviFSE2019} reported that flakiness due to coverage instrumentation is a common reason for failed coverage computation, which is manifested by performance failure or increased flakiness of non-deterministic tests. Tengeri et al. \cite{tengeriSANER16} reported cases where coverage instrumentation changes the behaviour of tests and their results. Finally, Dietrich et al. \cite{dietrichflaky} used instrumentation to intercept network errors to control flakiness caused by dependency on network connectivity.

\section{Background}
Instrumentation, the process of transparently adding functionality to a program, is often used in dynamic program analyses. For instance, to capture the runtime properties of a program. Instrumentation in Java often uses bytecode manipulation, facilitated by the availability of bytecode engineering libraries like \textit{asm} and \textit{javassist}. In addition, the Java Virtual Machine (JVM) supports instrumentation directly through agents, which can be deployed statically (via a JVM argument) or attached dynamically.

Instrumentation can cause interference in the behaviour of a program. By interference, we mean that execution is affected by the instrumentation of the program's code. Unavoidably, instrumentation does interfere with available resources as additional instructions use CPU/memory. Some cases of interference include: race conditions caused by timing issues introduced by instrumentation  \cite{ivankoviFSE2019}; instrumentation interfering with shared resources; classpath issues where the agent uses a different version of classes that are also part of the classpath of the program and its tests\footnote{In practice, those issues are often avoided by using dependency shading when building agents}.

Interference caused by instrumentation can change the outcome of tests, resulting in test flakiness. We use the following definition of test flakiness here: \textit{tests having different results on multiple executions for the same version of program code; or transitions of test outcomes for the same test across runs}. Here is an example that illustrates flakiness introduced by telemetry instrumentation from MockServer\footnote{\url{https://www.mock-server.com/}}. MockServer features functionality to mock HTTP/HTTPS, REST or RPC services for testing purposes. The test runs MockServer as a JUnit 5 extension, configured with the annotation \texttt{@MockServerSettings} to use port 8888 for the mocked service. The same port is used by the OpenTelemetry\footnote{\url{https://opentelemetry.io/}} APM collector service. When the test is executed with OpenTelemetry instrumentation, it fails with the exception that the port is already in use. A dependency on local network resources causes this flakiness.

\begin{lstlisting}[
    basicstyle=\ttfamily \scriptsize]
@ExtendWith({
    MockServerExtension.class,
    TestLoggerExtension.class,
})
@MockServerSettings(ports = {8787, 8888})
class MockServerExtensionConstructorInjectionWithSettings ...
\end{lstlisting}


\section{Study design}

This section lists the set of programs used in our study, the types of instrumentation used in the experiments, and the evaluation metrics we use. Our experiment requires rerunning each program's tests 20 times with the baseline (without instrumentation) and then with the use of each of the five instrumentations listed in Table~\ref{tab:instr}  (total of six). 

\subsection{Dataset}
We have used the set of programs used in Cordeiro et. al's study on manifesting flakiness by adding noise to the environment \cite{cord2021ASE}. In addition, they are 11 GitHub projects written in Java and use Maven for build automation. For two of the programs (\textit{ozone} and \textit{hbase}), this was not possible due to their size, as they take  considerably longer than the other programs ($>$30min per run). Thus, we have included only nine projects in our analysis (Table \ref{tab:programs}).

\begin{table}[]
\centering
\caption{Programs with GitHub repository name and commit ID/tag}
\label{tab:programs}
\begin{tabular}{l|l}
\hline

\textbf{repository} & \textbf{commit ID/tag}\\
\hline
OpenHFT/Chronicle-Queue & bec195b \\
CorfuDB/CorfuDB & b99ecff \\
Azure/azure-iot-sdk-java & a9226a5 \\
soabase/exhibitor & d345d2d \\
vaadin/flow & 6.0.6 \\
intuit/karate & 09bc49e \\
mock-server/mockserver & b1093ef \\
RipMeApp/ripme & 19ea20d \\
killbill/killbill & killbill-0.22.21\\
\hline

\end{tabular}
\end{table}

\subsection{Instrumentation}
While it is possible to craft an instrumentation that interferes with a particular test execution, we were interested in studying real-world instrumentation scenarios. To achieve this, we have identified several popular agents used for different purposes -- coverage capture, monitoring and profiling.

The instrumentation tools used in the experiment are listed in Table~\ref{tab:instr}. This includes Elastic APM, an application performance monitoring system, OpenTelemetry, JaCoCo for Java code coverage, IntelliJ's code coverage, and Java Flight Recorder, which collects profiling data for Java applications.

\begin{table}[]
\centering
\caption{Types of instrumentation}
\label{tab:instr}
\resizebox{\columnwidth}{!}{

\begin{tabular}{l|l|l|l}
\hline
\textbf{inst.} & \textbf{type} & \textbf{version}  &\textbf{rep}\\
\hline
elastic & APM & 1.34.1 & \url{/elastic/apm}\\
opentelemetry & APM & 1.12.0 & \url{/open-telemetry/opentelemetry-java}\\
jacoco & coverage & 0.8.8 & \url{/jacoco/jacoco/} \\
intellij & coverage & 1.0.684  & \url{/JetBrains/intellij-coverage} \\
jfr & profiling & N/A & \url{https://docs.oracle.com/en/java/java-components}\\
\hline
\end{tabular}}
\end{table}

\subsection{Evaluation Metrics}

\subsubsection{Flaky test count}

Flaky test count measures the number of tests over $N$ runs that result in different states across runs for a given instrumentation configuration. If the set of outcomes (as in JUnit) are $\{success, failure, error, skip\}$, a test, $t$, is flaky across runs (or configurations) if for any two runs, $r_1$ and $r_2$, there is a transition across test states, i.e. $r_1(t) \neq r_2(t) $ if we consider test runs as a mapping from tests to states.

\subsubsection{Flakiness score}
\label{sec:flaky-score}
The flakiness score measures the variability between test runs. For instance, we may already observe flakiness if a test fails only in 1/20 runs. It is still interesting to see whether we can see this changing to a higher (or lower) value with instrumentation being used, such as 5/20.

For each configuration (baseline or one particular instrumentation) for a program in the dataset, we compute this as follows:

\begin{itemize}
\item $FT$ is the set of tests that are flaky in all configurations. Note that this may not include tests that are flaky across configurations. Including those tests in the baseline would be problematic as adding another configuration may affect the flakiness scores for existing configurations.
\item For each run $i$ compute a set, $r_i$ consisting of pairs $(t, state) \in FT \times \{pass,fail,error,skip\}$
\item For each pair of runs, $r_i$ and $r_j$, compute the Jaccard distance, $d (r_i, r_j) = 1-\dfrac{|r_i \cap r_j |}{|r_i \cup r_j|}$, for $N$ runs.
 This yields $N(N-1)/2$ values (20 runs resulting in 190 values)
\item The distances across runs can then be $aggregated$ (e.g. mean or median) to obtain a flakiness score.

\end{itemize}

If the flakiness score is 0, there is no flakiness.
The flakiness score is greater than 0 if there is flakiness. For example, the baseline for a program's test results has some flakiness (one test fails once), but this one test fails a few more times with instrumentation. In this case, the flakiness score would go up even though the flaky count would remain the same.

\subsection{Experimental setup and process}

We run each program's tests 20 times in six different configurations (120 runs for each program),  i.e., baseline and the listed five types of instrumentation. Experiments were run on a computer with a 3.2 GHz 6-Core Intel Core i7 CPU with Oracle's Java SE Development Kit 8u301. We made the data from the experiments available online \url{http://www.bitbucket.org/unshorn/inst-study}.

\section{Results and Discussion}

\subsection{Experimental Results}
Detailed results for the experiments are shown below. Table~\ref{tab:results} shows the results for the experiments indicating flakiness counts. This includes the count of test outcomes (success, failure/error and flaky) and test runtimes (rt.) for each configuration (i.e., instrumentation). The numbers are averages for 20 runs.  Table~\ref{tab:scores} shows the flakiness scores discussed in Section \ref{sec:flaky-score} (configurations with zero values are omitted for brevity).

\begin{table}[]
\caption{Test results for all configurations, the numbers reported are the tests resulting in success, failure,  error or  skipped. Tests are reported across \textbf{20 test runs}. Numbers in brackets means variation across runs and report the mean. The number of  flaky tests across those runs are reported as well.  Runtimes are in seconds (rt(s)).}
\label{tab:results}
\centering
 \resizebox{\columnwidth}{!}{
	\begin{tabular}{|l|r|r|r|r|r|r|}
		\hline
		\textbf{program} & \textbf{config} & \textbf{succ.} & \textbf{fail/err} & \textbf{skip} & \textbf{flak.} & \textbf{rt.} \\
		\hline
\multirow{6}{*}{\textit{ChronicleQueue}}&baseline & 664 & 1 & 28 & 0 & (179.2) \\
 & opentele & (663.9) & (1.1) & 28 & 1 & (178.8) \\
 & jacoco & 664 & 1 & 28 & 0 & (251.9) \\
 & intelli & 664 & 1 & 28 & 0 & (215.2) \\
 & jfr & 664 & 1 & 28 & 0 & (178.8) \\
 & elastic & 663 & 2 & 28 & 9 & (190.0) \\
		\hline

\multirow{6}{*}{\textit{CorfuDB}}&baseline & (771.4) & (24.6) & 5 & 15 & (2784.2) \\
 & opentele & (768.7) & (27.2) & 5 & 38 & (3018.5) \\
 & jacoco & (773.0) & (23.0) & 5 & 10 & (2697.5) \\
 & intelli & (773.1) & (22.8) & 5 & 12 & (2664.9) \\
 & jfr & (769.2) & (26.8) & 5 & 74 & (2632.1) \\
 & elastic & (765.2) & (30.8) & 5 & 75 & (2685.6) \\
		\hline

\multirow{6}{*}{\textit{azure-iot-sdk-java}}&baseline & (4302.2) & (0.8) & 1 & 17 & (102.3) \\
 & opentele & (4302.8) & (1.5) & 1 & 16 & (102.6) \\
 & jacoco & (4303.2) & 0 & 1 & 0 & (102.8) \\
 & intelli & 4,305 & 0 & 1 & 0 & (102.5) \\
 & jfr & (4304.9) & (0.0) & 1 & 1 & (103.1) \\
 & elastic & (4303.2) & (0.7) & 1 & 13 & (103.8) \\
		\hline

\multirow{6}{*}{\textit{exhibitor}}&baseline & 106 & 0 & 0 & 0 & (84.4) \\
 & opentele & 106 & 0 & 0 & 0 & (88.4) \\
 & jacoco & 106 & 0 & 0 & 0 & (86.1) \\
 & intelli & 106 & 0 & 0 & 0 & (85.1) \\
 & jfr & (105.9) & (0.1) & (0.1) & 1 & (84.4) \\
 & elastic & 106 & 0 & 0 & 0 & (84.9) \\
		\hline

\multirow{6}{*}{\textit{flow}}&baseline & (4390.5) & (0.5) & 9 & 1 & (250.2) \\
 & opentele & (4390.5) & (0.5) & 9 & 1 & (265.3) \\
 & jacoco & (4389.5) & (1.5) & 9 & 1 & (231.8) \\
 & intelli & (4389.4) & (1.5) & 9 & 2 & (239.3) \\
 & jfr & (4390.5) & (0.5) & 9 & 1 & (245.9) \\
 & elastic & (4389.5) & (1.5) & 9 & 1 & (250.1) \\
		\hline

\multirow{6}{*}{\textit{karate}}&baseline & 604 & 1 & 0 & 0 & (57.5) \\
 & opentele & 604 & 1 & 0 & 0 & (57.3) \\
 & jacoco & (603.9) & (1.1) & 0 & 1 & (53.3) \\
 & intelli & 604 & 1 & 0 & 0 & (54.1) \\
 & jfr & 604 & 1 & 0 & 0 & (53.8) \\
 & elastic & 604 & 1 & 0 & 0 & (53.9) \\
		\hline

\multirow{6}{*}{\textit{killbill}}&baseline & (1017.9) & (0.5) & (24.6) & 65 & (133.6) \\
 & opentele & (1017.0) & (0.5) & (25.4) & 65 & (124.2) \\
 & jacoco & (1029.8) & (0.4) & (12.9) & 60 & (100.4) \\
 & intelli & (1019.6) & (0.5) & (22.9) & 65 & (124.4) \\
 & jfr & (1010.2) & (0.5) & (32.4) & 65 & (140.1) \\
 & elastic & (1017.4) & (0.5) & (25.1) & 65 & (121.6) \\
		\hline

\multirow{6}{*}{\textit{mockserver}}&baseline & 2,870 & 0 & 0 & 0 & (92.2) \\
 & opentele & (2867.9) & (1.1) & 0 & 1 & (92.7) \\
 & jacoco & (2869.8) & (0.1) & 0 & 3 & (92.7) \\
 & intelli & 2,870 & 0 & 0 & 0 & (91.9) \\
 & jfr & 2,870 & 0 & 0 & 0 & (92.7) \\
 & elastic & (2869.9) & (0.1) & 0 & 1 & (92.8) \\
		\hline

\multirow{6}{*}{\textit{ripme}}&baseline & (87.2) & (72.8) & 67 & 6 & (382.6) \\
 & opentele & (86.8) & (73.2) & 67 & 10 & (378.0) \\
 & jacoco & (87.7) & (72.3) & 67 & 4 & (361.6) \\
 & intelli & (88) & (72) & 67 & 2 & (354.4) \\
 & jfr & (87.6) & (72.4) & 67 & 6 & (367.5) \\
 & elastic & (87.0) & (73.0) & 67 & 6 & (377.7) \\
		\hline

	\end{tabular}}
\end{table}
\begin{table}[]
\caption{Flakiness scores (mean) over 20 runs for each program/configuration (0 values omitted for brevity).}
\label{tab:scores}
\center
	\begin{tabular}{|l|l|r|}
		\hline
			 \textbf{program} & \textbf{config} & \textbf{flakiness-score} \\ \hline
\textit{ChronicleQueue}&opentele & (0.200)\\
 & elastic & (0.339)\\
 \hline
\textit{CorfuDB}&baseline & (0.475)\\
 & opentele & (0.227)\\
 & jacoco & (0.556)\\
 & intelli & (0.484)\\
 & jfr & (0.119)\\
 & elastic & (0.352)\\
 \hline
\textit{azure-iot-sdk-java}&baseline & (0.082)\\
 & opentele & (0.210)\\
 & jfr & (0.200)\\
 & elastic & (0.029)\\
 \hline
\textit{exhibitor}&jfr & (0.200)\\
 \hline
\textit{flow}&baseline & (0.533)\\
 & opentele & (0.467)\\
 & jacoco & (0.467)\\
 & intelli & (0.370)\\
 & jfr & (0.556)\\
 & elastic & (0.533)\\
 \hline
\textit{killbill}&baseline & (0.450)\\
 & opentele & (0.502)\\
 & jacoco & (0.199)\\
 & intelli & (0.435)\\
 & jfr & (0.623)\\
 & elastic & (0.570)\\
 \hline
\textit{mockserver}&jacoco & (0.200)\\
 & elastic & (0.200)\\
 \hline
\textit{ripme}&baseline & (0.518)\\
 & opentele & (0.402)\\
 & jacoco & (0.486)\\
 & intelli & (0.556)\\
 & jfr & (0.474)\\
 & elastic & (0.330)\\
 \hline
	\end{tabular}
\end{table}

   \begin{figure}[]
        \centering
        \includegraphics[width=\columnwidth]{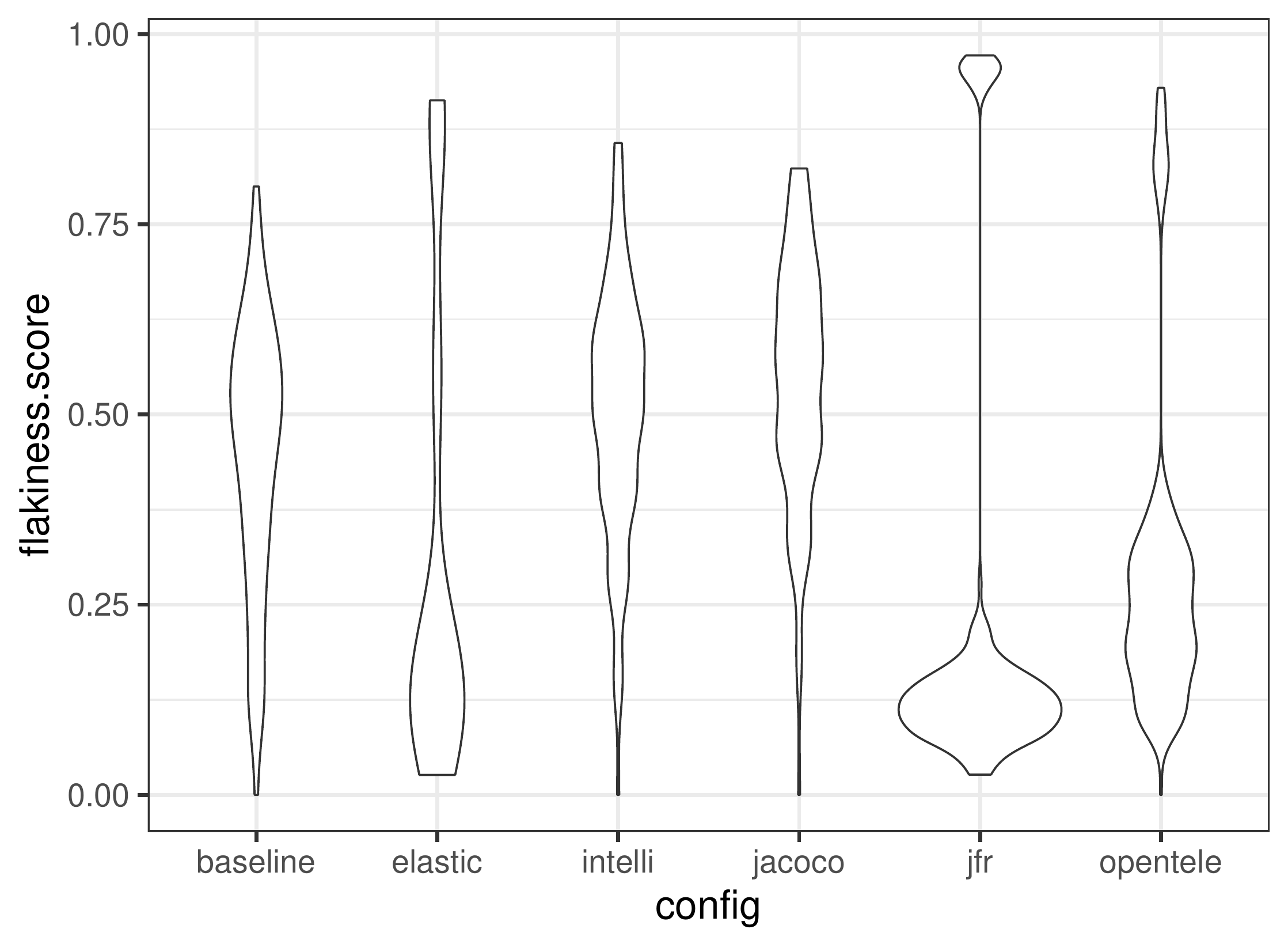}
        \caption{Variation of flakiness score for \textit{CorfuDB}}
\label{fig:boxplot}
    \end{figure}


In answering our research question if instrumentation changes the test outcome, we observe that, in general, there are no significant changes in test outcomes with or without the use of different instrumentation tools. However, there were cases in some of the programs we examined that showed variation in test outcomes. The two tests that are flaky across configurations, which are in the programs \textit{flow} and \textit{ripme} are due to external flakiness. The test,  \texttt{FrontendToolsLocatorTest::toolLocated}, in \textit{ripme} fails due to an error in executing in an external program, and \texttt{BaraagRipperTest::testRip} fails to access network resources, which is incidental and not caused by instrumentation. The only confirmed case of flakiness introduced by instrumentation (OpenTelemetry) is the failure of tests in \texttt{MockServerExtensionConstructorInjection\-WithSettingsMultiplePortTest} in \textit{mockserver} and \texttt{FrontendToolsLocatorTest::toolLocated} in \textit{flow}.

On the related question if instrumentation increases test flakiness over multiple runs; again, instrumentation does not appear to have a discernible effect on the stability of flaky tests. However, this is difficult to measure for most programs in the dataset as the count of flaky tests, seen in Table~\ref{tab:results}, is sparse for most configurations, with one flaky test per configuration being common. One of the programs where it shows noticeable results are for \textit{CorfuDB}. \textit{CorfuDB}, as listed in Table~\ref{tab:scores} has relatively more flaky tests when compared to the other programs. This varies from 10 to 75 flaky tests. As depicted in Figure~\ref{fig:boxplot} for \textit{CorfuDB}, flakiness across runs varies for the different instrumentations.

\subsection{Discussion of Findings}
Our experiment shows that while instrumentation may cause flakiness, this effect is rare. There are only a few cases, which show an impact of instrumentation on the presence of flaky tests. There is no unified pattern that we could identify with regard to specific instrumentation tools or configurations that would introduce/increase flakiness across the programs we study.

A related question is whether instrumentation can interfere with existing flakiness detection and prevention techniques. Given a large number of such techniques \cite{parry2021survey,rasheed2022test}, this discussion is incomplete.
We aim to investigate this question further in our future research.

For \textit{detection techniques based on static analysis} such as \cite{fatima2022flakify}, interference is possible (but may still be unlikely) simply because the instrumentation is not part of the analyses, and this may lead to additional false positives and/or false negatives. This effect can  be difficult to mitigate. 

Dynamic prevention and detection techniques such as \textit{RootFinder} \cite{wing2019rootcause} and \textit{saflate} \cite{dietrichflaky} (both use instrumentation)
can be prone to instrumentation order dependencies. While it is possible to craft examples showing this, it is unlikely to occur in practice. 

On the other hand, dynamic prevention and detection techniques that use a particular platform (such as \textit{NonDex} \cite{gyori2016nondex}  using a modified standard library or VMVM \cite{bell2014unit}) may also be sensible to instrumentation, as the instrumentation code itself is affected by those changes.

\section{Threats to Validity}
Even though the results of our experiments indicate the effects of instrumentation on test flakiness, there are threats to generalising them. First, the set of programs may not be representative, which we address using a set of programs from a previous flakiness study. As flaky tests can be non-deterministic and caused by environmental factors,  results may vary if the experiment is repeated. To account for such variance, we have executed the experiment 20 times and fixed the factors we control, such as the hardware, OS and JVM. Nonetheless, the results could change with more reruns of the experiment.

\section{Conclusion and Future Work}

In this paper, we discuss the possible impact of instrumentation on the presence of flakiness in test suites. We hypothesised that instrumentation does have an impact on the presence or frequency of flaky tests. To investigate this, we conducted an experiment using five Java instrumentation tools representing three different instrumentation types (APM, coverage and profiling). Our results from studying nine open-source programs show that instrumentation has little to no effect on the presence or frequency of flaky tests (i.e., it does not increase flakiness). These are preliminary results, and a more comprehensive study is needed. In the future, we plan to extend this study to include a more significant number of instrumentation tools and to study programs with a larger number of reruns.

\section*{Acknowledgment}
This work is funded by Science for Technological Innovation (SfTI) National Science Challenge (NSC) of New Zealand, grant number MAUX2004.
\bibliographystyle{IEEEtran}
\bibliography{bibliography}

\begin{thebibliography}{10}
\providecommand{\url}[1]{#1}
\csname url@samestyle\endcsname
\providecommand{\newblock}{\relax}
\providecommand{\bibinfo}[2]{#2}
\providecommand{\BIBentrySTDinterwordspacing}{\spaceskip=0pt\relax}
\providecommand{\BIBentryALTinterwordstretchfactor}{4}
\providecommand{\BIBentryALTinterwordspacing}{\spaceskip=\fontdimen2\font plus
\BIBentryALTinterwordstretchfactor\fontdimen3\font minus
  \fontdimen4\font\relax}
\providecommand{\BIBforeignlanguage}[2]{{%
\expandafter\ifx\csname l@#1\endcsname\relax
\typeout{** WARNING: IEEEtran.bst: No hyphenation pattern has been}%
\typeout{** loaded for the language `#1'. Using the pattern for}%
\typeout{** the default language instead.}%
\else
\language=\csname l@#1\endcsname
\fi
#2}}
\providecommand{\BIBdecl}{\relax}
\BIBdecl

\bibitem{parry2021survey}
O.~Parry, G.~M. Kapfhammer, M.~Hilton, and P.~McMinn, ``A survey of flaky
  tests,'' \emph{ACM Transactions on Software Engineering and Methodology
  (TOSEM)}, vol.~31, no.~1, pp. 1--74, 2021.

\bibitem{rasheed2022test}
S.~Rasheed, A.~Tahir, J.~Dietrich, N.~Hashemi, and L.~Zhang, ``Test flakiness'
  causes, detection, impact and responses: A multivocal review,'' \emph{arXiv
  preprint arXiv:2212.00908}, 2022.

\bibitem{hashemi2022empirical}
N.~Hashemi, A.~Tahir, and S.~Rasheed, ``An empirical study of flaky tests in
  javascript,'' in \emph{Proceedings of the 38th IEEE International Conference
  on Software Maintenance and Evolution (ICSME)}, 2022.

\bibitem{luo2014empirical}
Q.~Luo, F.~Hariri, L.~Eloussi, and D.~Marinov, ``An empirical analysis of flaky
  tests,'' in \emph{Proceedings of the 22nd ACM SIGSOFT international symposium
  on foundations of software engineering}, 2014, pp. 643--653.

\bibitem{costa2022test}
K.~Costa, R.~Ferreira, G.~Pinto, M.~d'Amorim, and B.~Miranda, ``Test flakiness
  across programming languages,'' \emph{IEEE Transactions on Software
  Engineering}, 2022.

\bibitem{ivankoviFSE2019}
M.~Ivankovi\'{c}, G.~Petrovi\'{c}, R.~Just, and G.~Fraser, ``Code coverage at
  google,'' in \emph{Proceedings of the 2019 27th ACM Joint Meeting on European
  Software Engineering Conference and Symposium on the Foundations of Software
  Engineering}, 2019.

\bibitem{wing2019rootcause}
W.~Lam, P.~Godefroid, S.~Nath, A.~Santhiar, and S.~Thummalapenta, ``Root
  causing flaky tests in a large-scale industrial setting,'' in
  \emph{Proceedings of the 28th ACM SIGSOFT International Symposium on Software
  Testing and Analysis}, 2019.

\bibitem{bell2018deflaker}
J.~Bell, O.~Legunsen, M.~Hilton, L.~Eloussi, T.~Yung, and D.~Marinov,
  ``Deflaker: Automatically detecting flaky tests,'' in \emph{2018 IEEE/ACM
  40th International Conference on Software Engineering (ICSE)}.\hskip 1em plus
  0.5em minus 0.4em\relax IEEE, 2018, pp. 433--444.

\bibitem{dutta2020detecting}
S.~Dutta, A.~Shi, R.~Choudhary, Z.~Zhang, A.~Jain, and S.~Misailovic,
  ``Detecting flaky tests in probabilistic and machine learning applications,''
  in \emph{Proceedings of the 29th ACM SIGSOFT International Symposium on
  Software Testing and Analysis}, 2020, pp. 211--224.

\bibitem{silva2020shake}
D.~Silva, L.~Teixeira, and M.~d’Amorim, ``Shake it! detecting flaky tests
  caused by concurrency with shaker,'' in \emph{2020 IEEE International
  Conference on Software Maintenance and Evolution (ICSME)}.\hskip 1em plus
  0.5em minus 0.4em\relax IEEE, 2020, pp. 301--311.

\bibitem{shi2019ifixflakies}
A.~Shi, W.~Lam, R.~Oei, T.~Xie, and D.~Marinov, ``ifixflakies: A framework for
  automatically fixing order-dependent flaky tests,'' in \emph{Proceedings of
  the 2019 27th ACM Joint Meeting on European Software Engineering Conference
  and Symposium on the Foundations of Software Engineering}, 2019, pp.
  545--555.

\bibitem{alshammari2021flakeflagger}
A.~Alshammari, C.~Morris, M.~Hilton, and J.~Bell, ``Flakeflagger: Predicting
  flakiness without rerunning tests,'' in \emph{2021 IEEE/ACM 43rd
  International Conference on Software Engineering (ICSE)}.\hskip 1em plus
  0.5em minus 0.4em\relax IEEE, 2021, pp. 1572--1584.

\bibitem{fatima2022flakify}
S.~Fatima, T.~A. Ghaleb, and L.~Briand, ``Flakify: A black-box, language
  model-based predictor for flaky tests,'' \emph{IEEE Transactions on Software
  Engineering}, 2022.

\bibitem{tengeriSANER16}
D.~Tengeri, F.~Horváth, A.~Besz\'{e}des, T.~Gergely, and T.~Gyim\'{o}thy,
  ``Negative effects of bytecode instrumentation on java source code
  coverage,'' in \emph{2016 IEEE 23rd International Conference on Software
  Analysis, Evolution, and Reengineering (SANER)}, vol.~1, 2016, pp. 225--235.

\bibitem{dietrichflaky}
J.~Dietrich, S.~Rasheed, and A.~Tahir, ``Flaky test sanitisation via on-the-fly
  assumption inference for tests with network dependencies,'' in \emph{2022
  22nd International Working Conference on Source Code Analysis and
  Manipulation (SCAM)}, 2022.

\bibitem{cord2021ASE}
M.~Cordeiro, D.~Silva, L.~Teixeira, B.~Miranda, and M.~d’Amorim, ``Shaker: a
  tool for detecting more flaky tests faster,'' in \emph{2021 36th IEEE/ACM
  International Conference on Automated Software Engineering (ASE)}, 2021.

\bibitem{gyori2016nondex}
A.~Gyori, B.~Lambeth, A.~Shi, O.~Legunsen, and D.~Marinov, ``Nondex: A tool for
  detecting and debugging wrong assumptions on java api specifications,'' in
  \emph{Proceedings of the 2016 24th ACM SIGSOFT International Symposium on
  Foundations of Software Engineering}, 2016.

\bibitem{bell2014unit}
J.~Bell and G.~Kaiser, ``Unit test virtualization with vmvm,'' in
  \emph{Proceedings of the 36th International Conference on Software
  Engineering}, 2014, pp. 550--561.

\end{thebibliography}

\end{document}